\documentclass[a4paper, 12 pt]{article}
\usepackage{amsmath}
\usepackage{amsfonts}
\usepackage{amsthm}
\usepackage{amscd}
\usepackage{amssymb}
\usepackage{graphicx}
\usepackage{color}
\usepackage{graphics}

\font\fontr=msbm10 scaled 1400
\newcommand{ \dst} {\displaystyle}

\newcommand{\la} {\lambda}

\newcommand{\eps} {\epsilon}

\newtheorem{thm}{Theorem}
\newtheorem{lem}{Lemma}

\newtheorem{prop}{Proposition}

\theoremstyle{definition}

\theoremstyle{remark}

\newcommand{\lt} {\ {\bf <}}
\newcommand{\rt} {{\bf >}\ }

\newcommand{\lb} {\left(}
\newcommand{\rb} {\right)}
\newcommand{\lbr} {\left\{}
\newcommand{\rbr} {\right\}}

\newcommand{\rd} {\right.}

\newcommand{\al} {\alpha}

\newcommand{\ga} {\gamma}
\newcommand{\bga}{\begin{array}{l}}
\newcommand{\ena}{\end{array}}
\newcommand{\bge}{\begin{equation}}
\newcommand{\ene}{\end{equation}}

\newcommand{\R} {\mbox{\fontr{R}}}

\def\comment#1{}

\def\withcomments{
\addtolength{\oddsidemargin}{-0.5 in}
\addtolength{\evensidemargin}{-0.5 in}
\newcounter{mycommentcounter}
\def\comment##1{\refstepcounter{mycommentcounter}%
  \ifhmode%
  \unskip%
  {\dimen1=\baselineskip \divide\dimen1 by 2 %
    \raise\dimen1\llap{\tiny -\themycommentcounter-}}\fi%
  \marginpar{\renewcommand{\baselinestretch}{0.8}%
    \footnotesize [\themycommentcounter]: \raggedright ##1}}
}

\withcomments

\begin{document}

\title{
Random integral currents. }
\author{ M. Zyskin}
\date{}
\maketitle

\begin{abstract}
For nice functions, invariant means over integral currents (certain generalized surfaces), can be uniquely defined. That may have applications to define Nambu-like string theory.
\end{abstract}

\section{Currents (generalized surfaces). }

Let $K$ be a compact set in $\R^n$.
Let $\Omega^m$ be a space of  $C^\infty$ differential forms on $K$, with $C^\infty$ norm. Space of $m$ currents $T_m$ is the space of continuous linear functionals on $\Omega^m$. Polyhedral chains are particular cases of currents, linear functionals on forms they define are integrals over polyhedral chains.

For $\omega \in  \Omega^m$,
$$
\bga
M (\omega) = \dst\mathop{\sup}_{x \in K} \| \omega (x) \|,
\\
\mbox{where for } \xi \in \Lambda^m, \| \xi  \| = \dst\mathop{\sup}_{\vert \ga \vert \leq 1} \lb  \xi \cdot \ga \rb, \quad \ga \mbox{ a simple m-vector. }
\ena
$$
Let $M(T)$ be the dual of $M (\omega)$,
$$
M(T) =\dst \mathop{\sup}_{\omega \in \Omega^m, M (\omega) \leq 1} \lb  T (\omega) \rb
$$
Space of normal currents $N$ is a linear space of currents with $M (T)+ M (\partial T) < \infty . $ 

Rectifiable currents $R$ are currents which may be approximated in $M$ semi-norm by integer Lipschitz chains, images under Lipschitz maps of polyhedral chains with integer coefficients.

Integral currents $I$ are normal currents such that $T$ and $\partial T$ are rectifiable currents. Integral currents form an abelian group. We equip integral currents
with flat semi-norm $\| \|_F$
$$
\|T \|_F = \inf \lb M (R) + M(S)  \rb , \quad T = R +\partial S, \quad \mbox{ R, S are rectifiable }
$$
It is clear that $\|T \|_F \leq M(T) $.
\section{Addition-invariant measure on integral currents $I_m$}
Let $f(X), X \in I_m$ be a bounded uniformly continuous function on space of integral $m-$currents $I_m$ with flat semi-norm.
Let
\bge
O_{f} = \lbr  f \lb X+Y \rb \vert Y \in I_m \rbr
\label{O_f}
\ene
be its $I_m$ orbit.

\begin{lem} A sequence in $O_f$ has a subsequence convergent point-wise in $I_m$ to a bounded continuous function on $I_m$.
\end{lem}
Let $\dst {B}^m_{\Lambda}  = \lbr T \in I_m \vert M(T) + M (\partial T) \leq \Lambda \rbr .$  $\dst {B}^m_{\Lambda} $ is compact in the flat semi-norm $\| \|_F$ \cite{f}. Using diagonal argument, it follows that  a sequence in  $O_f$  has  a  point-wise convergent
subsequence  on $I_m$. Indeed ,
let $n_1(k)$ be a subsequence uniformly convergent on $\dst {B}^m_{\Lambda}$, $n_2(k)$ a subsequence of $n_1 $ uniformly convergent on $\dst {B}^m_{2 \Lambda}$, $\ldots$ , $n_p(k)$ a subsequence of $n_{p-1} (k)$ uniformly convergent on $\dst {B}^m_{p \Lambda}$, $\ldots$.  Let   $  {\hat{k}} \equiv  {n_k (k )}.$
Then   $ f_{\hat{k}}(X) \rightarrow h(X)  $ point-wise,  $X\in I_m$. \qed
\vspace{5mm}

\begin{lem}
1) The orbit $O_f$ is relatively compact in the weak topology.\\
2) Weakly closed convex hull of $O_f$ is weakly compact
\end{lem}
A space dual to the space of continuous bounded functions on a normal topological space $S$ is the space $B$ of regular Borel measures on the field of closed sets,  and with norm being total variation.
$I_m $ is a space with a semi-norm $\|\|_F$ , and it is normal.
A sequence in $O_f$ has a point-wise convergent subsequence $ f_{\hat{k}}(X)$, and  $ f_{\hat{k}}(X)$ is uniformly bounded .
By dominated convergence theorem, for any measure $\mu \in B$ , $\dst\int f_{\hat{k}}(X)  d\mu $  is convergent .
Therefore the orbit  $O_f$  is relatively
sequentially
compact in the weak topology.
\\
2) follows from 1) , see \cite{schw} V 6.4  \qed
\vspace{5mm}

\begin{thm}Let $f(X), X \in I_m$ be a bounded uniformly continuous function on space of integral $m-$currents $I_m$ with flat semi-norm.
There is unique  mean  of $f (X)$ over $X\in  I_m$, invariant under addition in $I_m$. That is , there is uniquely defined constant $\lt f \rt$ ,  $\lt f \lb \cdot +Y  \rb \rt = \lt f \lb \cdot    \rb \rt$, such that for any $ \eps >0$ there exists $\lbr \la_i\in \R,  Y_i \in I_m  \left\vert \la_i\geq 0, \dst\mathop{\sum}_{i=1}^N \la_i =1 \rd \rbr,$  such that
\bge
\dst\mathop{\sup}_X \left\vert   \mathop{\sum}_{i=1}^N  \la_i f \lb X+Y_i \rb - \   \lt f \rt   \right\vert < \eps
\ene
\end{thm}
$I_m$ is an abelian group and  acts on continuous bounded functions by shifts; such action is distal.
From
Markov-Kakutani theorem \cite{schw}, \cite{ryl},  there is a unique fixed point of the action of $I_m$ on weakly compact convex hull of the orbit $O_f$. \qed
\vspace{5mm}

An easy  modification of the above argument can be used to compute mean over currents with prescribed boundary, by averaging over currents with zero boundary:
\begin{thm}
Let $\dst I_m^0 $ be space of integral m-currents $T$  with zero boundary , $\partial T =0$.  
Let $f(X), X \in I_m^0$ be a bounded uniformly continuous function.
There is unique  mean  of $f (X)$ over currents in $I_m^0$, invariant under addition in $I_m^0$.
\end{thm}
\vspace{5mm}

Motivated by applications, we give an example of a family of functions for which an invariant mean can be defined:
\begin{prop}
\end{prop}
Let $k$ be a $C^\infty$ 2-form on $\R^n$ with compact support, and with
$\max \lbr \| k\|,  \| d k\|\rbr < \infty $.
Let
\bge
\dst
g_{k} (X) = \exp \lb i  \int k \lfloor X  \rb \exp \lb  i \| X \|_{\mbox{F}}   \rb , X\in I_2 .
\ene
(where $\int k \lfloor X$ is an integral of a 2-form $k$ over integral current $X\in I_2$).  Let $G_{k }$ be the $I_2$ orbit of $g_{k } (X)$,
\bge
\bga
G_{k } =  \dst \lbr  g_{k } (X+Y) \vert X ,  Y \in I_2   \rbr  .
\ena
\ene

Functions in $G_k$ are uniformly bounded,
and equicontinuous, therefore the $I_2$ mean $\lt g_{k } \rt $ can be uniquely defined.

Indeed,  $\vert g_{k }  \vert \leq 1 $ , and 
$$
\bga
\vert g_{k} (X+Y) - g_{k} (\tilde{X} +Y) \vert =
\\
\left\vert \exp \lb i  \int k \lfloor \lb X+Y   \rb \rb \exp \lb i \| X+Y  \|_{\mbox{F}}   \rb
\lb 1- \exp \lb i  \int k \lfloor \lb X -\tilde{X}    \rb \rb \exp \lb i \| \tilde{X}+Y  \|_{\mbox{F}}   -  i \| X+Y  \|_{\mbox{F}}    \rb    \rb \right\vert
\\
\leq \vert \int  k \lfloor \lb X -\tilde{X} \rb \vert +       \| X - \tilde{X}  \|_{\mbox{F}}
\\
\leq \lb  1 +  \max \lbr \|k\|, \|dk \| \rbr \rb  \| X -\tilde{X}  \|_{\mbox{F}}
\ena
$$
(we used that $\dst \left\vert 1-e^{i \al} \right\vert \leq \left\vert \al  \right\vert , \al \in \R$).


\end{document}